\documentclass{aa}
\usepackage{psfig,graphics}
\begin{document}

   \thesaurus{%11 Galaxies
               (11.08.1; % Galaxies: halos,
                11.09.1; % Galaxies: individual,                     
                11.09.4; % Galaxies: ISM,
                11.11.1; % Galaxies: kinematics & dynamics
                11.13.2; % Galaxies:  magn. fields,
%               11.19.2; % Galaxies: spiral,
                11.19.6)} % Galaxies: structure.

   \title{The thermal and non-thermal gaseous halo of
NGC\,5775\thanks{Based on observations collected at the European Southern
Observatory, Paranal (Chile); Proposal N$\degr$: 63.N-0573(A)}}

   \author{R. T\"ullmann\inst{1,2} \and
          R.-J. Dettmar\inst{2} \and
          M. Soida\inst{3} \and
          M. Urbanik\inst{3} \and 
          J. Rossa\inst{2}}

   \offprints{rtuellma@eso.org}

   \institute{European Southern Observatory, Karl-Schwarzschild-Str.2,
        D-85748 Garching, Germany \and
        Astronomisches Institut, Ruhr-Universit\"at Bochum, 
        D-44780 Bochum, Germany \and
        Astronomical Observatory, Jagiellonian University
        PL-30-244 Krakow, Poland             
        } 

   \date{Received 14. August 2000; accepted 02. October 2000}

   \maketitle

   \begin{abstract}
In this letter we present first results from  spectroscopic observations
of Diffuse Ionized Gas (DIG) in the halo of NGC\,5775  obtained
with FORS1$^{1}$ attached to UT1/Antu of the Very
Large Telescope (VLT). At our slit position perpendicular to the disk 
(41\arcsec \ SE of the nucleus) the  emission of [\ion{N}{ii}]$\lambda$6583, 
[\ion{O}{iii}]$\lambda$5007, and [\ion{O}{ii}]$\lambda$3727 is detected 
out to 9\,kpc into the halo, allowing possible ionization 
mechanisms of the DIG to be examined. Photoionization models which
assume a dilute radiation field are able to fit the data for the disk,
but they cannot account for the line ratios measured
in the halo 
(e.g., [\ion{O}{i}]/${\rm H\alpha}$ or [\ion{O}{ii}]/${\rm
H\alpha}$). In particular they fail to predict the observed increase of
[\ion{O}{iii}]$\lambda$5007/${\rm H\alpha}$ with increasing $|z|$.
The most striking result concerns the kinematics of the halo gas. Velocities 
at high galactic latitudes drop from the midplane value to 
reach  the systemic velocity at $z \approx$\ 9\,kpc.
An analysis of VLA archive data of the polarized radio-continuum emission 
at 4.86\,GHz and 1.49\,GHz reveals that magnetic fields in the halo 
have a strong component perpendicular to the disk and are 
aligned with the H$\alpha$ and radio-continuum spurs in the halo. 
This can result either from a strong wind action or, more likely,
from the generation of dipolar magnetic fields. We briefly discuss
the interrelation of the magnetic field structure and gas dynamics,
in particular the role of magnetic fields in gas outflows, as well as
the possible heating of DIG by magnetic reconnection.

   \end{abstract}

%
%______________________________________________________________

\section{Introduction}
There is increasing evidence that halos of normal spiral galaxies contain
a complex multiphase interstellar medium (ISM) (for a review see,
e.g., Dahlem \cite{dahlemr}). Besides a hot component as
observed by X-ray emission ($>$10$^6$\,K) or UV absorption 
($\sim$0.5$\times$10$^5$\,K),
Diffuse Ionized Gas (DIG) constitutes another important phase of
typically 10$^4$\,K (Dettmar \cite{dettmar92}). This component is
easily observed by its $\ion{H}{ii}$ region like
line emission and is found to be widespread in galaxies with actively star 
forming disks (Rand \cite{rand96}, Dettmar \cite{dettmar99}, 
Rossa \& Dettmar \cite{rossa}).   

In the last couple of years two important problems have been discussed
with regard to the physics of this interstellar gas
phase. The observed line ratios are 
not explained in a straight forward way by pure photoionization models and
require  additional heating mechansims
of the halo gas (e.g., Golla et al. \cite{gollaetal}, Rand \cite{rand97}, 
T\"ullmann\,\&\,Dettmar \cite{tude}) while first
kinematic observations hint at a slower rotation of halo gas if compared 
to the underlying disk (Rand \cite{rand98}). 
Both kinds of observations could help to constrain models of the origin
of halo gas which in turn are important for the understanding of
galaxy 
evolution since large scale mass flows are of influence for star formation and
chemical evolution.

Only for a few objects, such as NGC\,891 (Dahlem et al. \cite{dahlem94}), 
NGC\,4631 (Golla\,\&\,Hummel \cite{gh4631}), and NGC\,5775, the
various components of the halo ISM are observed with
sufficient sensitivity and resolution to study possible physical processes
at their interfaces such as turbulent mixing or thermal conduction. 
For such processes yet another ISM component can be of importance, namely the
non-thermal component consisting of magnetic fields and  cosmic rays as
observed by its radio-continuum radiation. Any large scale gas motion and in 
particular its $z$-component
will also be of importance for dynamo theories which try to explain
 the large scale distribution of magnetic fields in galaxies. However, various
approaches to describe galactic dynamos are still very controversial and
more detailed information on galactic gas flows in halos as well as
magnetic field structures could help constraining such models.

For all these questions, the case of the Sc edge-on galaxy NGC\,5775
has received some attention recently, since marked
radio-continuum (Duric et al. \cite{duric}) and 
$\ion{H}{i}$ (Irwin \cite{irwin})
halo components are accompanied by a very extended DIG
 halo (Dettmar \cite{dettmar92},  Collins et al. \cite{coletal}) for
which a strong decline of the rotational velocity with height above
the plane ($z$) has been found (Rand \cite{rand00}).

In this letter we report on first spectroscopic results for the DIG halo
of NGC\,5775 with the VLT, confirming the strong $z$-dependence of the 
halo rotation presented by Rand (\cite{rand00}), along with a first detailed
determination of the magnetic field 
structure in the halo from VLA radio-continuum observations.

%  
%______________________________________________________________
\section{Observations}
\begin{figure}[!t]
\psfig{file=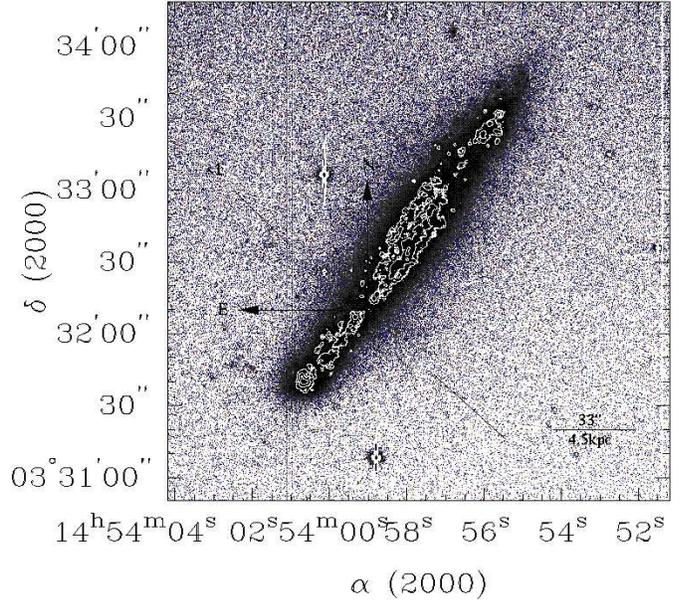,width=8.8cm,height=8cm,angle=0}
\caption{Continuum subtracted H$\alpha$ image of NGC\,5775 obtained
with EMMI at the NTT. H$\alpha$-contours shown for the disk reveal
 that s1 covers a gap of H$\alpha$ emission in the disk, cutting also 
through the NE and SW filament. %Slit s2 cuts the northern DIG parallel to
%the disk at a distance of 13\arcsec\ or 1.8\,kpc.    
}
\label{f1}
\end{figure}

Spectroscopic observations were carried out with FORS1\footnote{FOcal
 Reducer/low dispersion Spectrograph} at VLT/UT1 in longslit
 mode on 1999 June 14-15 at a median seeing of $0\farcs 7$. 
A Tek\,$2048 \times 2046$ CCD with a pixel size of $24\times24
{\rm \mu m}$ and a spatial resolution of $0\farcs2$ pix$^{-1}$ was
used for the measurements.
Grism GRIS\_600B+12 was chosen for the blue and GRIS\_600R+14
for the red spectral region, yielding a wavelength
coverage of 3500--5900\AA\ and 5250--7400\AA, respectively.
The slit length of FORS1 is $6\farcm 8$  and 
the selected slit width of $1\farcs 31$ gives an average resolution of 5.9\AA\
for the red and 6.2\AA\ for the blue grism, respectively. Our slit covers
large filamentary structures of H$\alpha$ emission in the NE and SW
 halo (Fig.~\ref{f1}).
For each grism the total integration time amounted to 3 hours, reaching
a noise level of 
$1.26\times10^{-19}$\,ergs cm$^{-2}$\,s$^{-1}$\,\AA$^{-1}$\,pix$^{-1}$.
Data reduction was performed in the usual manner using standard IRAF
software while the analysis was done within the MIDAS
environment. 

Maps of total power and the linearly polarized radio-continuum emission of
NGC\,5775 at 1.49\,GHz and 4.86\,GHz have been obtained from archive data
of  the Very Large Array
(VLA) of the National Radio Astronomy Observatory (NRAO\footnote{ NRAO 
is a facility of National Science Foundation operated under cooperative
agreement by Associated Universities, Inc.}). 
All data were recorded during 1989 -- 1994. 
The total observation time used  on 
source was about 10~hours at 4.86\,GHz (D-array), 14\,hours at
1.49\,GHz with C-array, and 2.5\,hours with D-array.
Data reduction has been performed with the AIPS software package
following standard procedures. 
The r.m.s. noise obtained in the final maps is 16\,$\mu$Jy/b.a. and 
100\,$\mu$Jy/b.a. in total power, and 7\,$\mu$Jy/b.a. and
25\,$\mu$Jy/b.a. in polarized intensity at 4.86\,GHz and 1.49\,GHz, 
respectively.

%  
%______________________________________________________________

\section{Results and discussions}
\subsection{Ionization}
\begin{figure*}[!pt]
\hspace{1.2cm}
\psfig{file=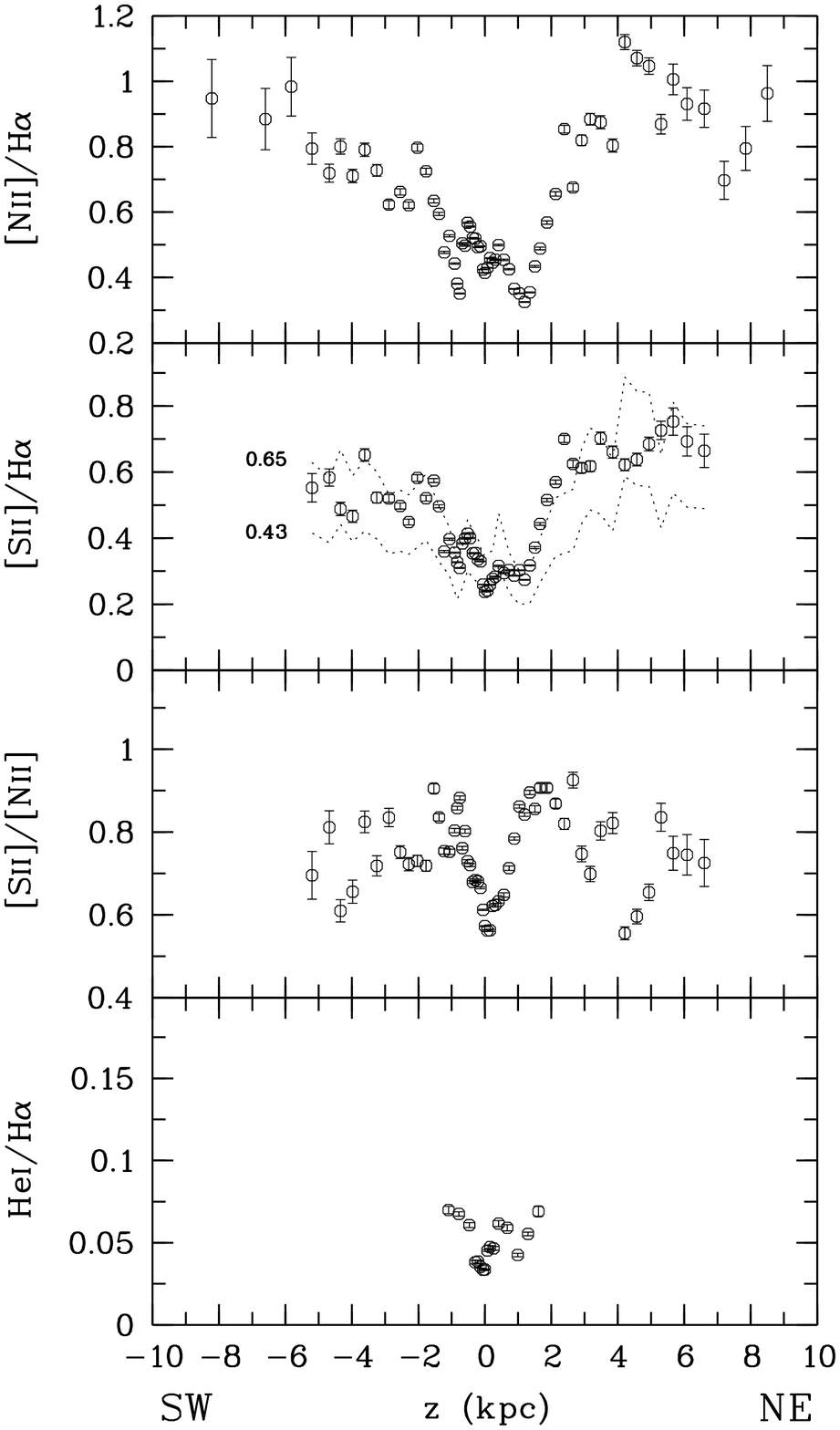,width=7.08cm,height=8cm}
\hspace{1.15cm}
\psfig{file=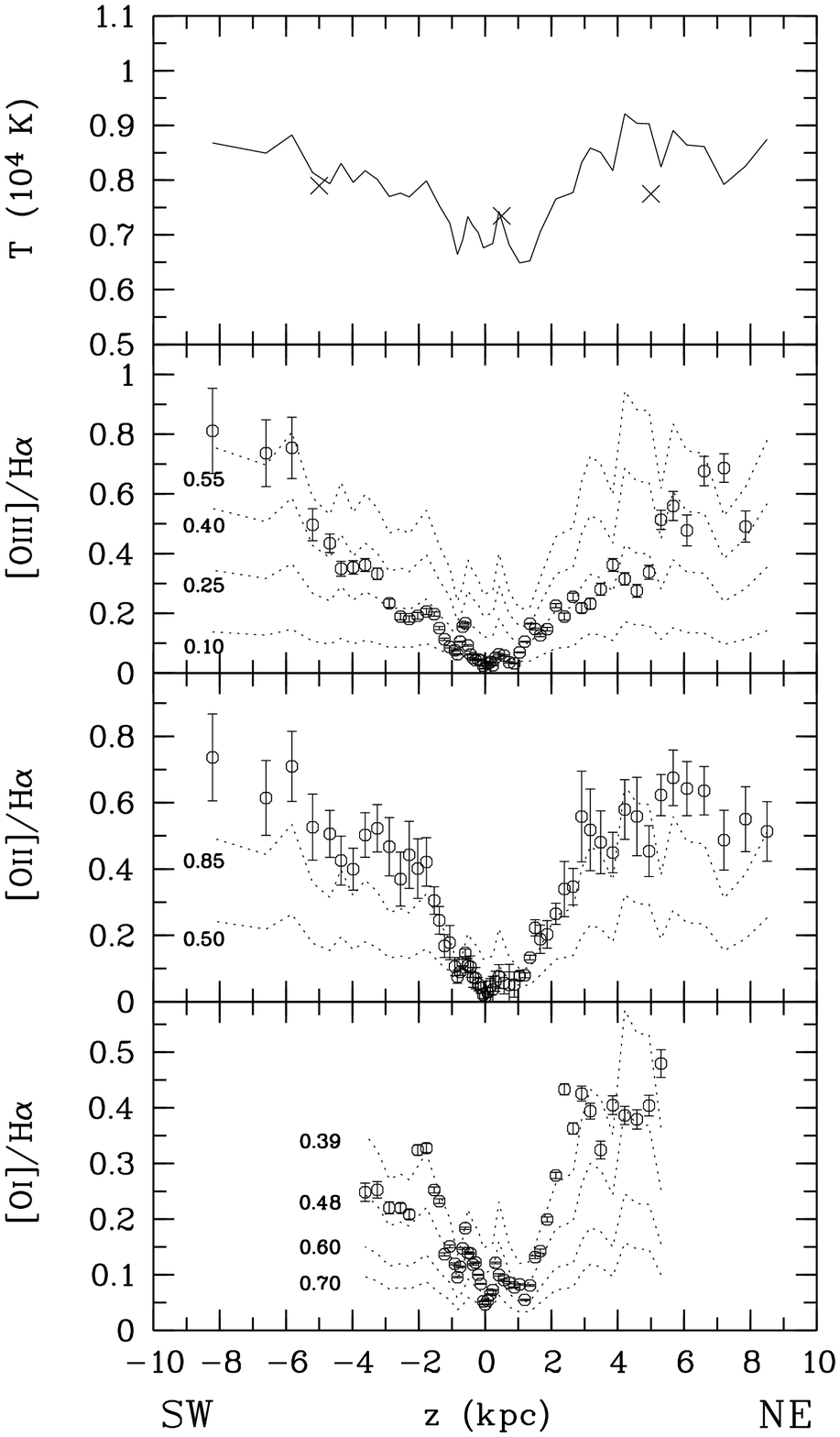,width=7.08cm,height=8cm}
\caption{Line ratios (corrected for interstellar extinction) plotted as a 
function of $|z|$. 
Theoretical line ratios (dotted lines) based on temperature profiles
(measured from nitrogen emission, upper right) are fitted to the data
by variing the corresponding ionization fraction. ``x"-symbols are from
temperature determinations using upper limits of $[\ion{N}{ii}]\lambda$5755
emission and are given as a consistency check.
}
\label{f2}
\end{figure*}

Several strong emission lines such as $[\ion{N}{ii}]\lambda$6583 or
$[\ion{O}{ii}]\lambda$3727 can be traced out to 66\arcsec \ or 9\,kpc 
(assuming a distance of D = 26.7\,Mpc) above the disk in both
directions, much further than
any previous imaging result. Similar findings have been reported 
by Rand \cite{rand00} for the western part of NGC\,5775. 
All measured line ratios are presented in Fig.~\ref{f2}. 
Some ratios, such as [\ion{S}{ii}] and [\ion{O}{i}], are lower in the
SW section, most likely due to a more patchy structure of the
filament. With an inclination of $i=84\degr$, NGC\,5775 is close to edge-on
and inclination effects are considered to be negligible. 
In order to clarify the issue of possible ionization mechanisms of the
DIG we compare our observational results with predictions of
pure photoionization models from Mathis (\cite{mathis}, Ma86
hereafter) and Domg\"orgen\,\&\,Mathis (\cite{doma}, DM94 hereafter).
If we adopt photoionization by OB stars as the only ionization
source and that radiation gets more and more dilute as it extends
towards the halo, these models should reproduce observational data, such
as the rising [\ion{O}{iii}]/H$\alpha$, non-linear changes in 
[\ion{S}{ii}]/[\ion{N}{ii}], or extreme ratios of 
[\ion{O}{i}]/H$\alpha$ and \ion{He}{i}/H$\alpha$.

[\ion{N}{ii}]/H$\alpha$ is fitted only for the disk by the model of Ma86
assuming a very soft radiation field ($\log q=-6$) and O7 stars with
effective temperatures
of 35000\,K as ionizing sources, whereas the DM94 model fails
completely. By no means DM94 can reproduce halo ratios for 
[\ion{S}{ii}]/H$\alpha$ larger than 0.66, while Ma86 gives correct
values ranging from 0.26 for the disk to 0.88 for the halo. 
Since these models predict an increase of [\ion{S}{ii}]/[\ion{N}{ii}]
towards the halo, most likely a consequence of slightly different
ionization potentials of S$^{+}$ and N$^{+}$, they are unable to explain the
non-linear trend visible in Fig.~\ref{f2}. 
Another important ratio is \ion{He}{i}/H$\alpha$ which could be
measured only close to the disk-halo interface of NGC\,5775
($-$1.1\,kpc\,$<\hspace{-0.05cm}z\hspace{-0.05cm}<$\,1.65\,kpc),
making this detection nevertheless unique concerning its $z$ extent.
However, both models underestimate this ratio by a factor of two.
 
The most striking feature in Fig.~\ref{f2} is the rise of  
[\ion{O}{iii}]/H$\alpha$ with increasing $z$ which is not
explained by pure photoionization codes. Generally this ratio
declines from values of about 2 in the disk to 0.8 in the
halo. 

Both models also fail to predict [\ion{O}{ii}]/H$\alpha$. The lowest
predicted values vary around 1, whereas the observed data never
exceed this limit. 
[\ion{O}{i}]/H$\alpha$ reveals with 0.33 for the SW and 0.48 for the
NE halo by far the highest values ever measured for the DIG. Ma86 and
DM94 are able to fit ratios only up to 0.1. 

The extreme line ratios of \ion{He}{i}/H$\alpha$ ($>$\,0.05) and 
[\ion{O}{i}]/H$\alpha$ ($>$\,0.3), the shapes of [\ion{O}{iii}]/H$\alpha$
and [\ion{S}{ii}]/[\ion{N}{ii}] together with an unprecedented
DIG extent of 9\,kpc (see also Rand \cite{rand00}) make this galaxy
the most outstanding and challenging testbench for future ionization models.

Since common photoionization models are unable to reproduce the data for
halo DIG in external galaxies as well as the Reynolds-layer of
the Milky Way, recent studies (e.g., Rand \cite{rand98}, Reynolds et al. 
\cite{rehatu}, T\"ullmann \& Dettmar \cite{tude}) have pointed out the
need to involve additional
ionization and/or heating mechanisms such as shocks, photoelectric
heating by dust, or magnetic reconnection. These mechanisms should
increase the electron
temperature without affecting the ionization stage of an atom and
dominate in the halo over collisional ionization.  
Following Haffner et al. (\cite{haretu}) and
Collins\,\&\,Rand (\cite{coll00}, CR hereafter) we used the 
$[\ion{N}{ii}]\lambda$6583 emission line to determine a temperature 
profile (shown in the upper right diagram of 
Fig.~\ref{f2}) which varies between 6500 and 9200\,K.
This profile can be used to derive theoretical line ratios 
(dotted lines in Fig.~\ref{f2}) for oxygen
and sulfur. In order to fit the observed ratios, the
ionization fraction as the only free parameter is changed 
accordingly. 
For consistency we also derived electron temperatures for the eastern part of 
NGC\,5775 which are based on upper limits of the
$[\ion{N}{ii}]\lambda$5755 emission line (Osterbrock, 
\cite{oster}). They have been estimated to be 7350\,K in the disk (at
0.5\,kpc), 7750\,K for the NE filament, and 7900\,K for the SW
filament (both measured at 5.0\,kpc), respectively.
These values are in good agreement with our temperature determinations
mentioned above and also with profiles obtained by CR for the central
and western part of this galaxy.

Furthermore, plots on the right of Fig.~\ref{f2} reveal prominent gradients
which decrease as the ionization stage of oxygen increases. A likely physical
reason could be a rising temperature towards the halo, according to
Reynolds et al. (\cite{rehatu}). This can simply
be shown if we plot theoretical oxygen line ratios per ionization fraction 
vs. electron temperature. As a result [\ion{O}{i}]/H$\alpha$
increases much faster with $T$ than 
[\ion{O}{ii}]/H$\alpha$ or [\ion{O}{iii}]/H$\alpha$. Hence, different
temperature dependencies of oxygen ionization stages are able to reflect the
observed gradients.

\subsection{Kinematics}
\begin{figure}[t]
\psfig{file=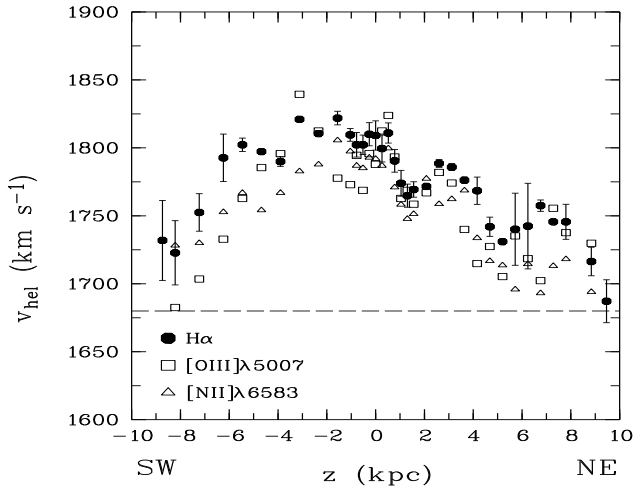,width=8.4cm,height=6.5cm}
\caption{Heliocentric velocities as a function of $z$ for the
minor slit s1. Dashed lines indicate the systemic velocity $v_{\rm sys}$ as
measured by Irwin (\cite{irwin}). 
No qualitative differences are visible between the kinematical
behavior of {\rm H$\alpha$}, $[\ion{O}{iii}]\lambda$5007,
 and $[\ion{N}{ii}]\lambda$6583.}

\label{f3}
\end{figure}

The most surprising result revealed by our deep spectra concerns
the kinematics of DIG at large $z$.
The bright emission lines of hydrogen, nitrogen, and oxygen have been 
used to derive
heliocentric velocities for the DIG as a function of $z$ (vertical distance
above the plane). %and $R$ (radial distance). 
As can be seen in
Fig.~\ref{f3} the gas velocity %at the minor slit s1 
is with $1822\pm 8$\,km\,s$^{-1}$ highest at $z$\,=\,0\,kpc and $-$1.6\,kpc,
decreasing slowly towards larger $z$ distances, 
finally reaching values close or
equal to the systemic velocity of 1681\,km s$^{-1}$ (Irwin
\cite{irwin}) at $z=9$\,kpc. 
The measured trend is explainable assuming the DIG not to rotate at high 
galactic $z$. A similar result was recently reported by Rand (\cite{rand00})
for two different slit positions.  
This allows us to examine whether the halo DIG rotates regularly since 
our slit position  and slit 1 of Rand
(\cite{rand00}) both have a distance of 45\arcsec\ from the
center corresponding to 6.1\,kpc but on opposite sides of the disk. The
maximum rotational velocities at these  
positions (with respect to $v_{\rm sys}$ at $z=$\ 2\,kpc) deviate only 
by 4\,km\,s$^{-1}$ and the dependence of $v_{\rm rot}(z)$ is to first order
in agreement with a projected regular velocity field. 
However, different gradients imply that the kinematics of the DIG halo is more complex than a symmetric
and static model would allow for. This requires more modelling including
the shape of the dark matter potential and possible influences of outflow
kinematics. The resolution of our spectra unfortunately is unsufficient to
check for the velocity dispersion of the lines. 

Since our spectra have  very high sensitivity at very good
spatial resolution, Fig.~\ref{f3} also indicates some 
small-scale structure 
in the velocity field with localized minima at $z=$ $-$8.5, $-$4.0, 1.5, 5.2, 
and 9.5\,kpc. Close to the disk, in particular in the NE, these changes can be
explained by dust obscuration, ``shifting" the peak velocity at the
center to $z$\,=\,$-1.6$\,kpc. 
An R-band image of NGC\,5775 gives evidence for  large amounts of dust,
mainly in the northern part, extending from the disk plane 
1.8\,kpc into the halo. The kinematical structures at larger $z$ are not
directly associated with observed dust features and hint at either
an inhomogeneous DIG distribution along the line of sight or effects from
flows.

\subsection{Magnetic field structure in the halo of NGC\,5775}

\begin{figure*}[!pht]
\psfig{file=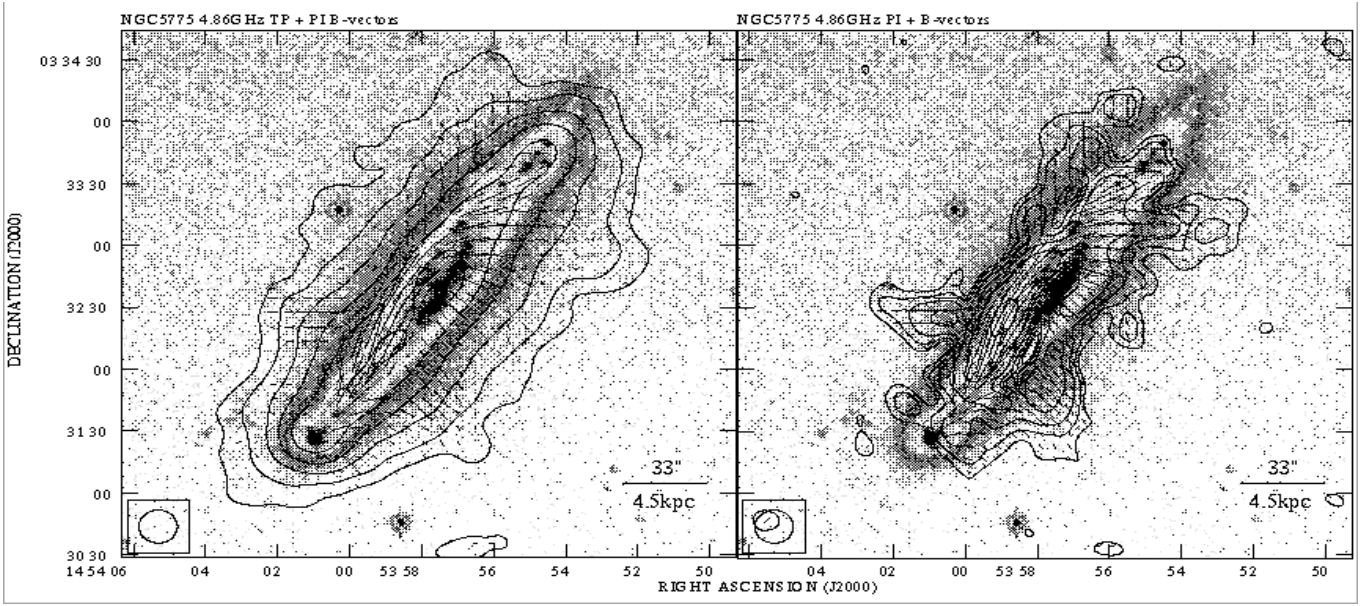,width=18cm,height=8cm,angle=0}
\caption{VLA radio-continuum maps with total power (left), polarized
 intensity (right) and resulting  B-vectors overlaid on a H$\alpha$ image of 
NGC\,5775. Contours are at  3, 8, 21, 55, 144, 377, and 
610$\times$16\,$\mu$Jy/beam for TP and 
3, 5, 8, 13, 21, and 34$\times$7\,$\mu$Jy/beam for PI,  for the B-vectors
a length of 1\arcsec \ corresponds to 10\,$\mu$Jy.
}

\label{f4}
\end{figure*}

In Fig.~\ref{f4} we present maps of total power, polarized intensity, and 
apparent polarization B-vectors
at 4.86\,GHz overlaid upon an enhanced H$\alpha$
image. NGC\,5775 shows an extended radio envelope which can
be traced beyond 2\,kpc from the disk plane. While close to
the disk plane the polarization B-vectors are generally disk
parallel, at heights above 1\,kpc they form a X-like
pattern, corresponding also to similarly shaped extensions
of polarized intensity.

Before discussing the magnetic field structure we make a
comparison of the polarization information at both
frequencies to estimate the possible influence of Faraday
rotation and depolarization. 
For heights less than 1\,kpc our estimates for a regular azimuthal magnetic
field oriented parallel to the disk, assuming pressure balance with 
cosmic rays and typical thermal electron densities of 0.03 cm$^{-1}$,
 imply that NGC\,5775 is Faraday thick at 4.86\,GHz close to the disk
plane.
Therefore B-vector orientations are not conclusive. However, a
reasonable agreement between the orientations of
polarization angles at 4.86\,GHz and 1.49\,GHz, a relatively
high polarization degree (reaching 30\%) at 4.86\,GHz, and
the depolarization at 1.49\,GHz by a factor of about 0.3\,--\,0.4
support the assumption that above 1\,kpc the halo of
NGC\,5775 is Faraday-thin down to 1.49\,GHz. Thus, if the B-vector
orientations at two frequencies differ there locally
by some tens of degrees, the rotation angle at 1.49\,GHz is
smaller than $90\degr$, hence the Faraday rotation bias at
4.86\,GHz does not exceed $10\degr$. However, in the SW part
of the halo (R.A.$_{2000}=14^{h} 53^{m} 54^{s}$
Dec$_{2000}=+03\degr 33\arcmin 15\arcsec$), similar
orientations of B-vectors at both frequencies are apparently
in conflict with the depolarization at 1.49\,GHz by a factor
of 0.1, which raises the suspicion that it might be Faraday-thick.
On the other hand numerical simulations show that
this depolarization can also be due to pure Faraday
dispersion in random fields, without substantial Faraday
rotation.

From the above comparison we conclude that the observed B-vectors
at 4.86\,GHz in the halo at heights $|z|\hspace{-
0.1cm}>$\hspace{-0.1cm} 1\,kpc are not significantly
affected by Faraday rotation and that therefore the magnetic
field in the halo has a significant vertical component. Such a
vertical magnetic field component strong enough to be seen
in emission can be generated in two ways. A spherically
blowing galactic wind could result in an enhanced vertical
component of the magnetic field. While the poloidal magnetic
field preserves its quadrupole symmetry, resulting in B-vectors
parallel to the plane below $z$\,$\simeq$\,1\,kpc,
characteristic X-shaped structures develop at larger heights
(Brandenburg et al. \cite{brand}). 

An alternative way to obtain the vertical B-vectors is the
generation of the dipolar (called A-type) poloidal magnetic
field produced by the dynamo process. The principal
condition is a rigid rotation extending over a substantial
range of galactocentric radius and a large vertical
scaleheight of the ionized gas, both being true for
NGC\,5775 (Lehnert \& Heckman \cite{lehe}). 
In this case the B-component parallel to the disk would vanish.
Unfortunatley, this way of distigushing both mechanisms cannot be used
here because of the strong Faraday effects close to the disk.
However, we note that the detailed analysis of Faraday rotation angles
reveals field reversals across the plane (excluding $z$\,$<$\,1\,kpc) 
in favour of a dipolar poloidal field.

Whether a vertical decrease of rotational speed, as seen in
Fig.~\ref{f3}, makes this
field mode growing faster remains yet to be investigated but
cannot be excluded. We note that the existing dynamo models
which account for rotation speed falling with $z$ yield very
different results. While little effect of vertical rotation
decrease upon the dynamo efficiency was obtained for a
classical dynamo concept (Brandenburg et al. 1993) a special
model of a supernova-driven dynamo by Ferri\`ere\,\&\,Schmitt
(\cite{fer}) fails in this case to generate any stable,
growing magnetic field mode. The dynamo models in case of a
vertical rotation drop-off can be validated if only more
information on the magnetic field structure in halos along
with kinematical information would be available.
\begin{figure}[!hpt]
\psfig{file=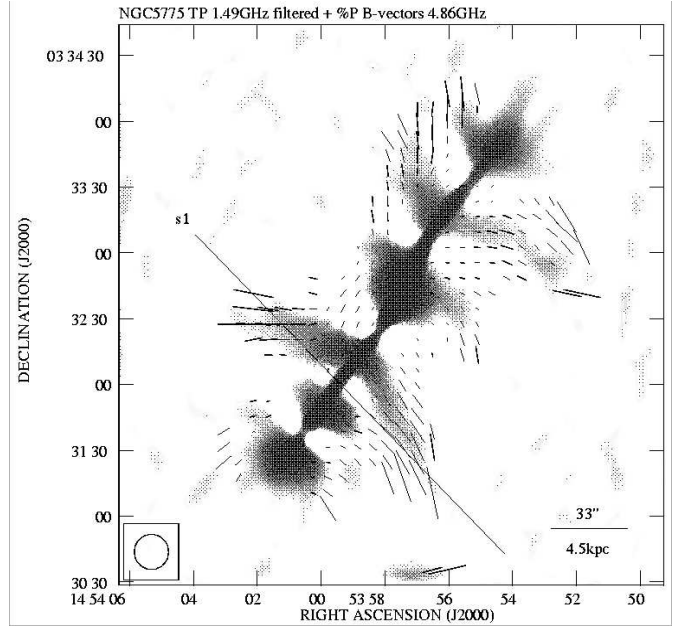,width=8.8cm,height=8.3cm,angle=0}
\caption{The strongly filtered total power image of
NGC\,5775 at 1.49\,GHz with a median filter applied. It
shows in detail the loci of brightness maxima in the
distribution of the total power brightness.}
\label{f5}
\end{figure}
 
However, as noted by Collins et al. (\cite{coletal}) we need
to consider the influence of the vertical magnetic field
structure on the cosmic ray propagation and the gas flows.
To clearly demonstrate the correlation of the radio-
continuum spurs with the magnetic field structure we have
applied a median filter to the total power map at 1.49\,GHz,
which revealed structural components inclined to the galaxy
plane in the same X-shaped manner as the polarization B-vectors
(Fig.~\ref{f5}). 

The polarization features  are anchored in the disk plane with
the SE one being attached to a local radio bright region in the
disk, most likely a star forming region. 
The total power spur at
R.A.$_{2000}=14^{h} 53^{m} 53\fs 1$, Dec$_{2000}=+03\degr
32\arcmin 55\arcsec$ curves towards south in its outer part;
the same behaviour is seen for the polarization vectors at
both frequencies. We suspect that they are associated with
cosmic ray electron streaming from intensively star-forming
regions along the inclined magnetic lines. The inclined
spurs extend down to the disk plane which allows us to
speculate that magnetic lines may do the same, as expected
for the A-mode. We also note that the cosmic ray propagation
along the magnetic lines of the regular field may yield an X-shaped
distribution of the polarized intensity
(Fig.~\ref{f4}). Similarly, an easier streaming of the
ionized gas along magnetic fields of the discussed geometry
can give rise to the occurrence of H$\alpha$-emitting spurs
coincident with the radio ones and to an increased vertical
scale of ionized gas in regions of magnetic structures
highly inclined to the disk plane. 
 
Finally we want to address the question whether the energy 
stored in the magnetic field could in principal help solving
the heating problem discussed in sect. 3.1. Magnetic
reconnection as a heat source for DIG  was suggested earlier, e.g.,
by Birk et al. (\cite{Birk}) or Reynolds et al. (\cite{rehatu}), the 
latter based on an analogy from solar physics proposed by  Raymond 
(\cite{raymond}). In the following we use the heating rate for
magnetic reconnection 
$G_{\rm mag}$ as derived by Lesch \& Bender (\cite{lesch}): 
%$G_{\rm mag}$=$\frac{B}{8\pi v_{\rm A}/L}$, 
$G_{\rm mag} = B^{2} v_{\rm A} / 8\pi L$, 
with $v_{\rm A}$ the Alf\' en
velocity and $L$ the dissipation length. 
For characteristic regions in the spurs the minimum total magnetic
field strength implied by the pressure balance with cosmic rays is 
 $\ge$\,5\,$\mu$G. With this value we obtain the required heating rates
as determined by Reynolds at al. (\cite{rehatu}) for DIG densities of
typically $n_{\rm e} =10^{-3}$\,cm$^{-3}$ if 
$L$ is in the order of parsec, not unreasonable for structures in the ISM.
Since this heating rate depends on the Alf\' en velocity and thus on
$n_{\rm e}^{-1/2}$ the increasing importance of the additional heating
source with decreasing particle density would be a natural consequence.
Reconnection could also be responsible for the growth of 
ordered magnetic field structures in the spurs. It has been suggested that
this process could help to reorganize the field structure in a
galactic fountain flow (e.g., Kahn \cite{kahn}). 

%
%______________________________________________________________

\section{Summary and conclusions}
The most important results can be summarized as follows:
\begin{itemize}
\item emission lines of DIG can be traced out to 9\,kpc in the
halo of NGC\,5775
\item the rotational velocity of the DIG drops from 
the rotational value at the midplane and approaches the
systemic velocity at distances $z>$\,6.0\,kpc
\item pure photoionization models are unable to reproduce the extreme
line ratios of [\ion{O}{i}]/${\rm H\alpha}$, \ion{He}{i}/${\rm
H\alpha}$, and [\ion{O}{iii}]/${\rm H\alpha}$
as measured for the halo 
\item a high degree of polarization and a strong vertical
component of the magnetic field are observed in the halo. We
found some, yet weak evidence for the dipolar magnetic
field
\item a correlation exists between extraplanar DIG and magnetic fields,
evidenced by the alignment of magnetic B-vectors with prominent halo
features in the DIG and radio-continuum distribution
\item we suggest that reconnection processes play a substantial role in both 
heating the DIG structures in the halo and
building up regular fields away from the disk plane

\end{itemize}

\begin{acknowledgements}        

RT acknowledges support through the ESO Studentship Programme.
Research in this field is supported through grant 50 OR 9707 from DLR at 
Ruhr-University and through grant no. PB4264/P03/99/17 from  Polish
Research Committee (KBN) at Jagiellonian University. The project
also benefitted from the exchange programme between both universities. 
We thank the referee for asking the right questions and RJD appreciates
valuable discussions with H. Lesch and A. Shukurov on the role of magnetic
fields for the physics of the ISM.
\end{acknowledgements}

\end{document}